\documentclass[twocolumn]{aastex631}

\shorttitle{GJ 1214 Non-detection}
\shortauthors{Spake et al.}
\graphicspath{{./}{figures/}}
\begin{document}

\title{Non-detection of He I in the atmosphere of GJ1214b with Keck/NIRSPEC, at a time of minimal telluric contamination}

\author[0000-0002-5547-3775]{Jessica J. Spake}
\affiliation{Division of Geological and Planetary Sciences, California Institute of Technology, Pasadena, CA 91125, USA}

\author[0000-0002-9584-6476]{A. {Oklop{\v{c}}i{\'c}}}
\affiliation{Anton Pannekoek Institute for Astronomy, University of Amsterdam, 1090 GE Amsterdam, Netherlands}

\author{L. A. Hillenbrand}
\affiliation{Department of Astronomy, California Institute of Technology, Pasadena CA 91125, USA}

\author[0000-0002-5375-4725]{Heather A. Knutson}
\affiliation{Division of Geological and Planetary Sciences, California Institute of Technology, Pasadena, CA 91125, USA}

\author[0000-0003-0534-6388]{David Kasper}
\affil{Department of Astronomy \& Astrophysics, University of Chicago, 5640 South Ellis Avenue, Chicago, IL 60637, USA}

\author[0000-0002-8958-0683]{Fei Dai}
\affiliation{Division of Geological and Planetary Sciences, California Institute of Technology, Pasadena, CA 91125, USA}

\author[0000-0003-2066-8959]{Jaume Orell-Miquel}
\affiliation{Instituto de Astrofísica de Canarias (IAC), 38205 La Laguna, Tenerife, Spain}
\affiliation{Departamento de Astrofísica, Universidad de La Laguna (ULL), 38206 La Laguna, Tenerife, Spain}

\author[0000-0003-2527-1475]{Shreyas Vissapragada}
\affiliation{Division of Geological and Planetary Sciences, California Institute of Technology, Pasadena, CA 91125, USA}

\author[0000-0002-0659-1783]{Michael Zhang}
\affiliation{Division of Geological and Planetary Sciences, California Institute of Technology, Pasadena, CA 91125, USA}

\author[0000-0003-4733-6532]{Jacob L.\ Bean}
\affil{Department of Astronomy \& Astrophysics, University of Chicago, 5640 South Ellis Avenue, Chicago, IL 60637, USA}

\begin{abstract}
Observations of helium in exoplanet atmospheres may reveal the presence of large gaseous envelopes, and indicate ongoing atmospheric escape. \cite{2022A&A...659A..55O} used CARMENES to measure a tentative detection of helium for the sub-Neptune GJ 1214b, with a peak excess absorption reaching over 2\% in transit depth at 10830 \AA. However, several non-detections of helium had previously been reported for GJ 1214b. One explanation for the discrepancy was contamination of the planetary signal by overlapping telluric 
absorption- and emission lines. We used Keck/NIRSPEC to observe another transit of GJ 1214b at 10830 \AA\, at a time of minimal contamination by telluric lines, and did not observe planetary helium absorption. Accounting for correlated noise in our measurement, we place an upper limit on the excess absorption size of 1.22\% (95\% confidence). We find that the discrepancy between the CARMENES and NIRSPEC observations is unlikely to be caused by using different instruments or stellar activity. It is currently unclear whether the difference is due to correlated noise in the observations, or variability in the planetary atmosphere.

\end{abstract}

\keywords{}


\section{Introduction} \label{sec:intro}
The compositions of the most common type of planet in the Galaxy - sub-Neptunes - remain unclear (e.g. \citealt{2021JGRE..12606639B}). Population studies suggest that the majority of sub-Neptunes are born with significant hydrogen-helium envelopes, which may be lost over time \citep{2013ApJ...775..105O, 2013ApJ...776....2L, 2017AJ....154..109F}. One way to observe atmospheric escape, and infer the presence of a hydrogen-helium atmosphere, is via transit observations covering the triplet of absorption lines of metastable helium at 10830 \AA\footnote{All absorption lines listed in this paper are at air wavelengths.} (e.g.  \citealt{OklopcicHirata2018, Spake2018}). However, the relatively small radii of sub-Neptunes can make such observations difficult. Additionally, some stellar host types are unfavourable for populating the metastable state of helium in exoplanet atmospheres. Both \cite{Oklopcic2019} and \cite{2022MNRAS.512.1751P} found that K-type stars are the most favourable.  Indeed, the majority of strong helium detections to date have been from large planets around K stars \citep{Allart2018, Nortmann2018, Salz2018, Kirk2020, paragas2021metastable}.

The smallest planets with helium detections so far have been TOI 560.01 \citep{2022AJ....163...67Z}, with a radius of 2.84 R$_{\earth}$, and  GJ1214b \citep{2022A&A...659A..55O}, with a radius of 2.74 R$_{\earth}$. Amongst all of the sub-Neptunes observed at 10830A, the TOI 560 system is one of the most favourable targets because of its young (500 Myr) and chromospherically active K4 host star \citep{2021arXiv211013069B}. For TOI 560.01, \cite{2022AJ....163...67Z} measured an excess absorption of 0.68\% in the core of the helium triplet at mid-transit. %, which corresponds to an extent of 2.5 planetary radii.  
\cite{2022A&A...659A..55O} observed GJ1214b, which orbits an older, quiet, M4 star (3-10 Gy, \citealt{2009Natur.462..891C}), and found a tentative 2.1\% transit depth in the core of the helium triplet. These detections were exciting for several reasons - the two planets are by far the smallest with evidence for helium absorption; they belong to the sub-Neptune class of exoplanets, whose bulk densities are consistent with a wide range of possible atmospheric compositions \citep[e.g.][]{2010ApJ...712..974R, zeng2019}; and the atmosphere of GJ1214b in particular has been extensively studied \citep[e.g.][]{2010Natur.468..669B, 2012ApJ...747...35B, 2014Natur.505...69K} but had not previously shown evidence of either atomic or molecular absorption.

The large helium signal observed by \cite{2022A&A...659A..55O} suggests that GJ1214b has a substantial H/He envelope. However, three non-detections of helium had previously been reported for GJ1214b, using IRTF/SpeX; VLT/X-Shooter; and Keck/NIRSPEC (\citealt{Crossfield_2019, 2020RNAAS...4..231P, 2020AJ....160..258K}).   To explain the discrepancy, \cite{2022A&A...659A..55O} showed that the three previous datasets were contaminated with telluric absorption and emission lines. The telluric lines overlap the helium line at the radial velocity of the planet as it accelerates through a transit, which may explain why \cite{2022A&A...659A..55O} were the only ones to detect the planetary signal.
We have observed an additional transit of GJ1214b with Keck/NIRSPEC, using the same setup as \cite{2020AJ....160..258K}. Our observations were fortuitously timed such that the planetary absorption signal was minimally contaminated by telluric absorption and emission lines. In Sections \ref{sec:obs} and \ref{sec:data-reduc} we describe our observations and data reduction, respectively. In Section \ref{sec:absspec} we describe how we constructed the planetary absorption spectrum, and in Section \ref{sec:modelling} we describe our atmospheric modelling. In Section \ref{sec:discussion} we discuss our results and compare them to previous observations, and we conclude in Section \ref{sec:conclusion}.

\section{Observations} \label{sec:obs}
To observe one transit of GJ1214b, we used the upgraded \citep{Martin2018} NIRSPEC spectrograph \citep{mclean1998,mclean2000} on the 10-meter Keck II telescope, for one half night, on 2019 May 18 UT (Semester 2019A, PI: Hillenbrand). Observations were performed using NIRSPEC in its echelle mode with the N1 (Y-band) filter, which has an effective wavelength coverage from 0.941 to 1.122 $\mu$m, with some overlap and no gaps between the orders. The helium triplet lines appear in both Order 70 and Order 71, at rest wavelengths of 10830.34 \AA, 10830.25 \AA, and 10829.09 \AA. We used both orders in our analysis.  We used the $0.576\arcsec  \times 12\arcsec$ slit, resulting in a spectral resolution of $R= 19,000$ (corresponding to $\Delta v \approx 16$~km~s$^{-1}$) at the wavelength of the helium line. 

We used a standard ABBA nodding pattern with 4$\arcsec$ throw between the nod positions and individual exposures of 210 s. Similarly to \cite{Kirk2020}, and \cite{2021AJ....162..284S}, we  %(accidentally) 
used the thin blocker filter which lowers background thermal emission. We obtained a sequence of 40 individual spectra, with SNR averaging around 35 per resolution element throughout the night. The mid-transit time coincided with the airmass minimum, which remained below 1.30 for the entire half-night. Before the bulk of our observations we also took a long exposure with 4 co-adds of 210 s each, in the `A' position. 

\section{Data reduction} \label{sec:data-reduc}
We followed the same data reduction procedure as that of \cite{2021AJ....162..284S}, barring minor adjustments listed here. In the previous work we analysed a similar NIRSPEC dataset for WASP-107b, and demonstrated that our pipeline produced a planetary absorption spectrum that agreed within the 1 $\sigma$ uncertainties to the Keck/NIRSPEC pipeline of \cite{Kirk2020}. For the current analysis, the main difference in our approach was an improvement in the correction of bad pixels which fell inside the spectrum trace. First, we made a night-averaged flat field by median-combining 81 five-second exposures of a halogen lamp, from which we subtracted the median of eleven five-second exposures taken with the lamp off, to account for detector bias and dark current. Then, we applied the iterative bad pixel algorithm from REDSPEC \citep{2015ascl.soft07017K}, which we had previously translated to Python and used in \cite{2021AJ....162..284S}, to the master flat field. We saved the position of the bad pixels found on the master flat to create a bad pixel mask. Next, we applied REDSPEC to each of the science frames, to correct for cosmic rays. 

To extract 1D spectra from the raw frames, we used a customized version of the NIRSPEC Data Reduction Pipeline\footnote{ \url{https://github.com/Keck-DataReductionPipelines/NIRSPEC-Data-Reduction-Pipeline}} (NSDRP), which is written in Python. The same code was used in \cite{2021AJ....162..284S}, barring one adjustment, described below. Before extraction, NSDRP performs flat-field correction; removal of telluric emission by pairwise subtraction of consecutive images; and spatial- and spectral rectification of the trace. In this work, we adapted the spectral extraction algorithm so that it replaced the value of the bad pixels from the bad pixel mask with the median of the spatial trace, scaled to fit the pixel column. We also used the average spectral trace to find 5-sigma outliers inside the spectral trace (presumably cosmic rays which REDSPEC was unable to correct), and replaced them in the same way. This process is similar to the Optimal Extraction algorithm described in \cite{1986PASP...98..609H}. To reduce the single, long, `A' exposure at the start of our observations, we subtracted the sum of the next 4 `B' exposures at the pairwise-subtraction stage.   

To determine the wavelength solution for each 1D spectrum, we followed a similar method to \cite{2020arXiv201202198Z}. First, we selected a model stellar spectrum from the PHOENIX grid appropriate for GJ 1214 (T$_\mathrm{eff}$=3000K; $log(g)$=5.0; [M/H]=+0.5; \citealt{2013A&A...553A...6H}) shifted into the star's rest frame, and added a telluric transmission spectrum\footnote{\url{https://www.gemini.edu/observing/telescopes-and-sites/sites}}.  We then down-sampled the resulting template spectrum to match the instrumental resolution (R=19,000).  We modelled the wavelength solution as a third-order Chebyshev polynomial (mathematically equivalent to monomials, but well-behaved around zero), and the stellar continuum as a fifth-order Chebyshev polynomial.  We then used SCIPY's differential evolution to minimize the $\chi^{2}$ between the continuum-normalised observations and the template spectrum.  This method resulted in continuum-normalized, wavelength-calibrated, 1D spectra in the observer's rest frame, with line positions aligning with the model to better than one pixel.

\begin{figure*}
    \centering
    \includegraphics[width=\textwidth]{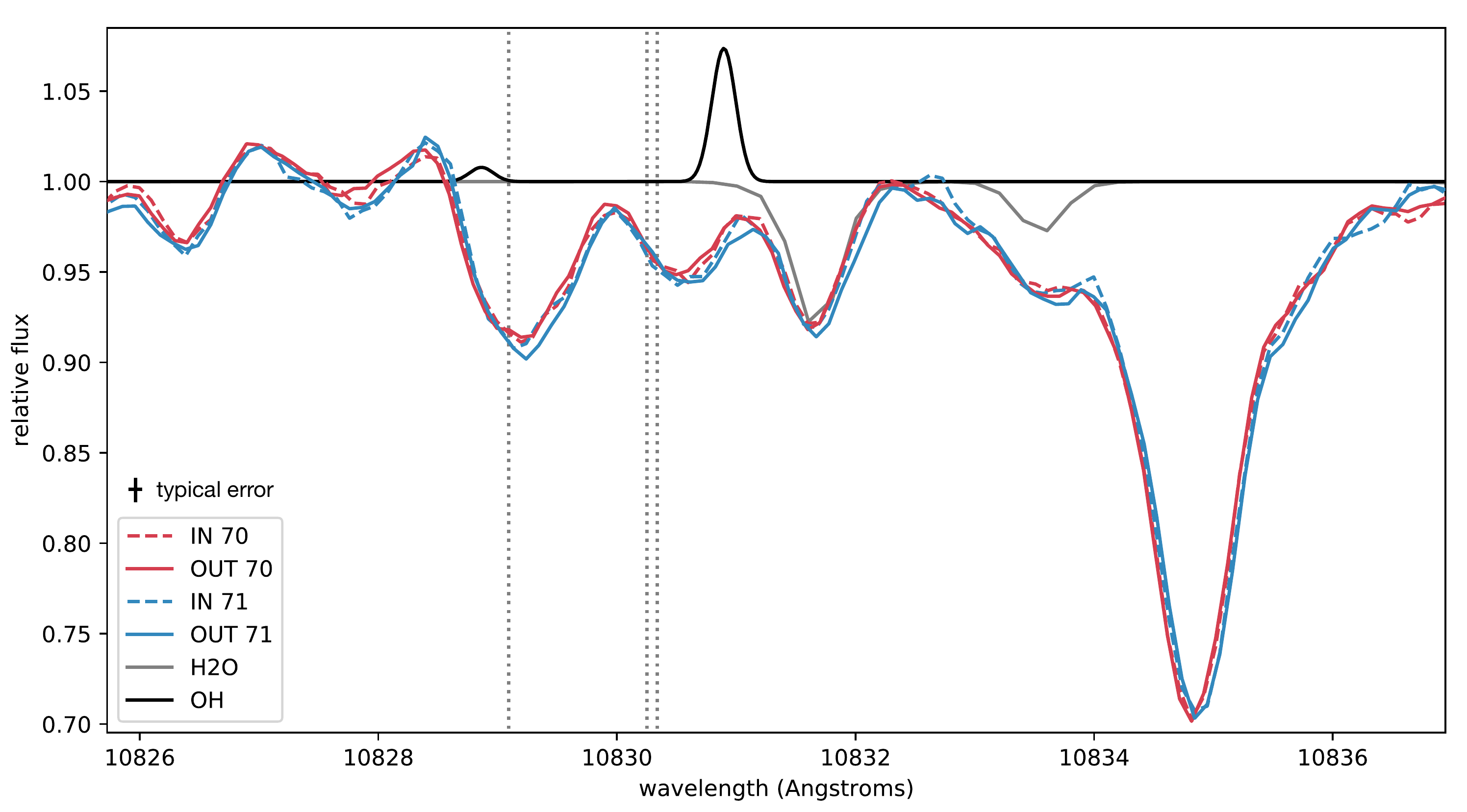}
    \caption{Comparison of in-transit and out-of-transit spectra of GJ1214b obtained with Keck/NIRSPEC, in the stellar rest frame.  Grey and black lines show the telluric absorption (H$_2$O) and emission (OH) lines, respectively. Vertical dotted lines indicate location of helium absorption line triplet.}
    \label{fig:invsout}
\end{figure*}

\section{Constructing the absorption spectrum}\label{sec:absspec}
We created an absorption spectrum for each order (70 and 71) separately, but followed the same method for both. First, we created an average stellar spectrum by taking the weighted mean of all spectra which fell entirely before the first or after the fourth transit contact. There were 8 pre-transit spectra (including the one long exposure at the start), 21 post-transit spectra, and 12 spectra between first and fourth contact.  Figure \ref{fig:invsout} shows the average `OUT' stellar spectrum and the average `IN' transit spectrum for each order, shifted into the stellar reference frame. In the same figure we have also shown the position of the telluric H$_{2}$O absorption and OH emission lines. 

\begin{figure*}
    \centering
    \includegraphics[width=0.7\textwidth]{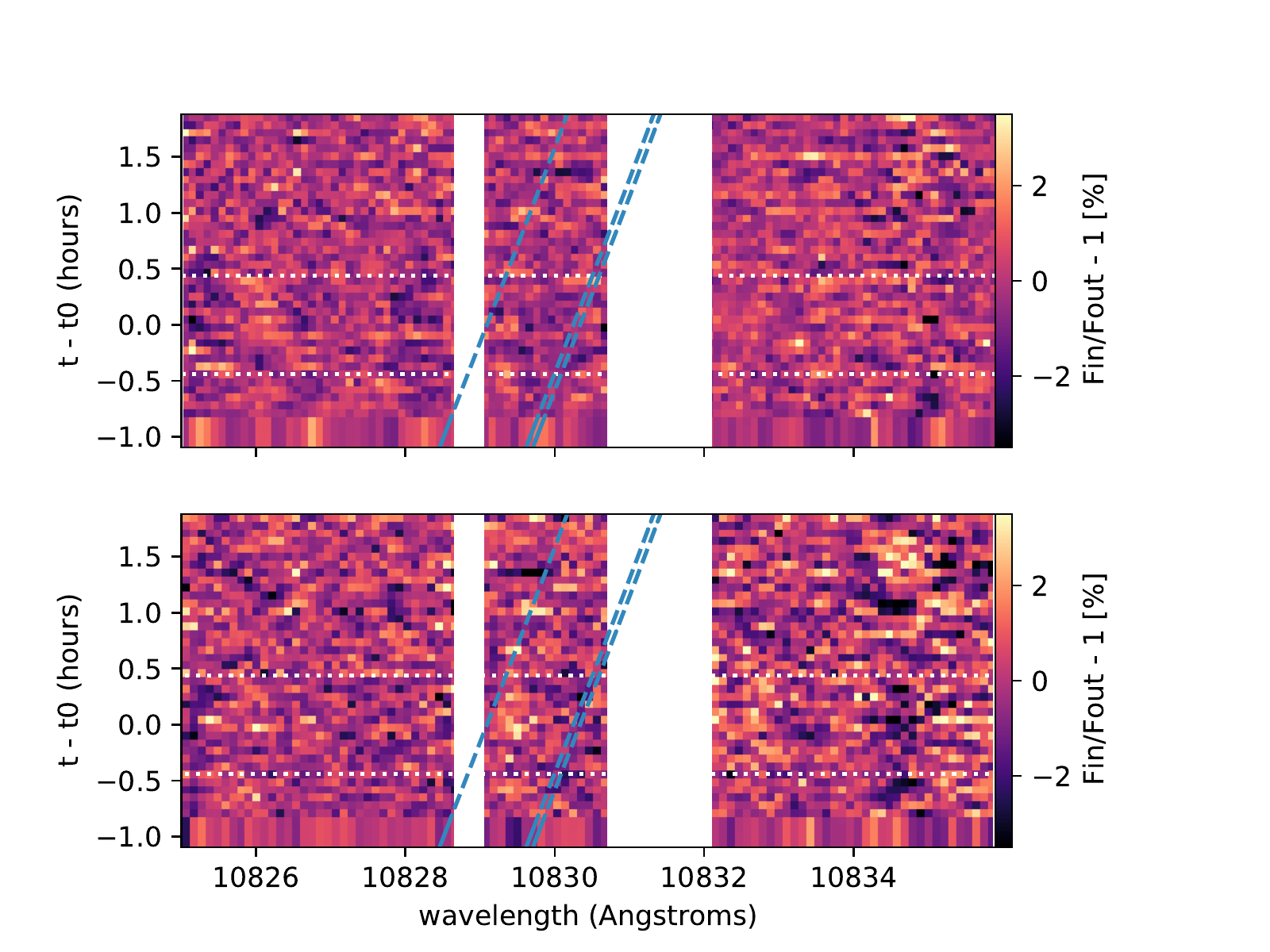}
    \caption{Individual planet absorption spectra through time. The top panel is order 70, and the bottom panel is order 71. The start and end of the transit are indicated by the dotted white lines. The radial velocity trace of the expected position of the helium triplet is indicated by dashed blue lines. Pixels contaminated by telluric absorption and emission (see Figure 1) have been removed from the analysis (white vertical bars).}
    \label{fig:2d}
\end{figure*}

Next, we divided each individual spectrum by the mean stellar spectrum and subtracted the result from 1 to find the excess absorption at each point in time. A 2D plot showing each of these excess absorption spectra is shown in Figure \ref{fig:2d}. The bottom row shows the first exposure, which is made of four co-adds, and is therefore wider than the rest. We set the extent of the color range from -3.5\% to +3.5\%, to approximately match the colour scale of Figure 2 of \cite{2022A&A...659A..55O} for ease of comparison. We did not correct the contamination from telluric absorption or emission lines, since any correction would be inherently imperfect and we aimed to be as conservative as possible in placing upper limits on the planetary signal. Instead, we followed the more conservative of the two procedures used by \cite{2022A&A...659A..55O}, and did not use any telluric-contaminated pixels in our analysis. Specifically, we masked two regions of  pixels which fell between 10828.65 - 10829.05 \AA\, and 10830.7 - 10832.1 \AA. These are the white regions in Figure \ref{fig:2d}.

\begin{figure*}
    \centering
    \includegraphics[width=\textwidth]{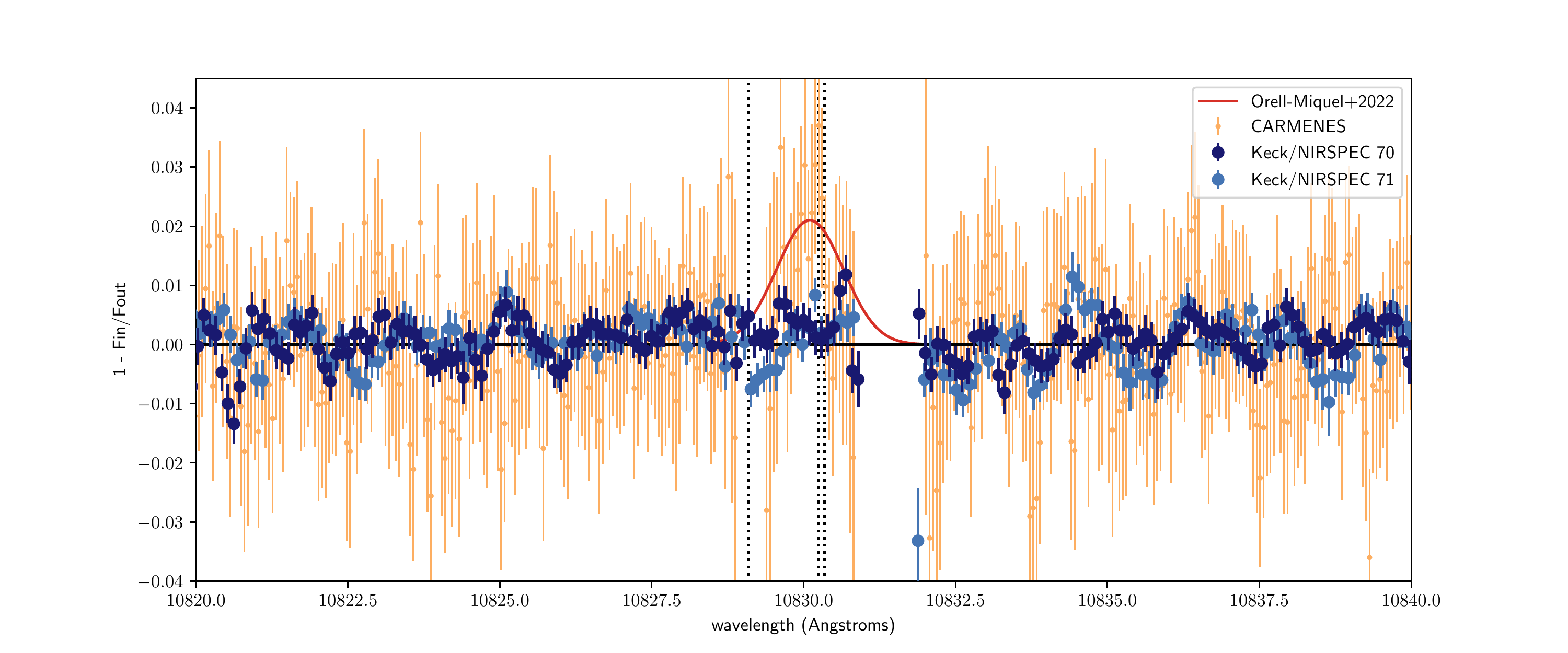}
    \caption{Comparison of the planetary absorption spectrum of GJ 1214b from CARMENES and Keck. Red line shows the best-fit Gaussian model from \cite{2022A&A...659A..55O}; black line is the fiducial mass-loss model based on \cite{Salz2016}. Note that pixels contaminated by telluric emission have been removed from the analysis.}
    \label{fig:comparison}
\end{figure*}

Next, we shifted each excess absorption spectrum onto the rest frame of the planet, re-sampled them at the wavelength solution of the first exposure, and summed to create a total planet absorption spectrum, shown in Figure \ref{fig:comparison}. No telluric-contaminated pixels were included in the sum. A gap in the data and large uncertainties near the contaminated regions can be seen in the final spectrum, because of the reduced number of data points used.  There is no obvious planetary absorption signal at the wavelength of the helium lines, but correlated noise on the order of 1\% is evident. It is currently unclear whether the correlated noise is instrumental or astrophysical. To estimate a conservative upper limit on the helium signal strength, accounting for correlated noise, we fit Gaussian profiles to the combined absorption spectrum from both orders, shifting the central wavelength along by one pixel-width each turn. We used only the overlapping regions of the two orders (covering approximately 10815 to 10845 \AA), resulting in 280 trials. We used uniform priors for the excess absorption between -3 and 3\% (meaning we allowed for emission signatures as well), and uniform priors on the standard deviation between 0.01 and 1.0 \AA. 
\cite{2022A&A...659A..55O} found an absorption signal blueshifted by approximately 0.17 \AA\, away from the centre of the two strongest helium lines.  In our observations, the largest-area Gaussian within 0.17 \AA\, either side of the two strongest lines had an amplitude of 0.38\%, and a standard deviation of 0.49 \AA. The standard deviation of the amplitudes of all 280 trials was 0.42\%, which we take as the systematic error. The 95\% upper limit is therefore 1.22\%, or 1.37 R$_{p}$.

Our absorption spectrum is consistent with that of \cite{Kasper_2020} within the 1 $\sigma$ uncertainties between 10827 and 10831 \AA. In Figure \ref{fig:comparison} we also show the absorption spectrum of GJ 1214b measured by CARMENES, from \cite{2022A&A...659A..55O}, which has been analysed using the same method (including masking out telluric-contaminated pixels).

\section{Atmospheric modelling}\label{sec:modelling}
We used the one-dimensional Parker-wind model of \cite{OklopcicHirata2018} to constrain the atmospheric properties of GJ 1214b. This model has been extensively used to interpret helium absorption spectra of exoplanets at 10830 \AA\, \citep[e.g.][]{Spake2018, KreidbergOklopcic2018, Mansfield2018, Kasper_2020, paragas2021metastable}. For a given stellar UV spectrum, planet mass loss rate, and thermospheric temperature, it calculates the radial density and velocity profiles of an escaping, isothermal, hydrogen/helium atmosphere, 
and calculates the excitation level populations.  For consistency, we used the publicly-available, model UV spectrum of the star GJ1214 from \cite{2022A&A...659A..55O}. The spectrum was calculated from coronal models fit to X-ray spectra from XMM-Newton observations \citep{2014ApJ...790L..11L}, and UV spectra from Hubble Space Telescope \citep{2016ApJ...824..101Y}, following the procedure of \cite{2011A&A...532A...6S}. We used the updated planet mass of 8.17 M$_{\earth}$ from \cite{2021AJ....162..174C} - which is 30\% greater than the previous measurement used by \cite{Kasper_2020} in their model grid. 

We followed the same parameter exploration scheme of \cite{Kasper_2020} to aid comparison with their non-detection. Using the 1D density profile of metastable helium obtained from the atmospheric models, and assuming a spherically symmetric planetary atmosphere, we computed a mid-transit absorption spectrum at wavelengths around 10830\AA\, for grids of model atmospheres with helium number fractions of 2\%, 5\%, 10\%, and 20\%. Each grid spans a broad range of values in thermospheric temperature (2,000 – 8,000 K) and mass-loss rates ($10^{7} – 10^{11}$ g/s$^{-1}$).

\begin{figure*}
    \centering
    \includegraphics[width=\textwidth]{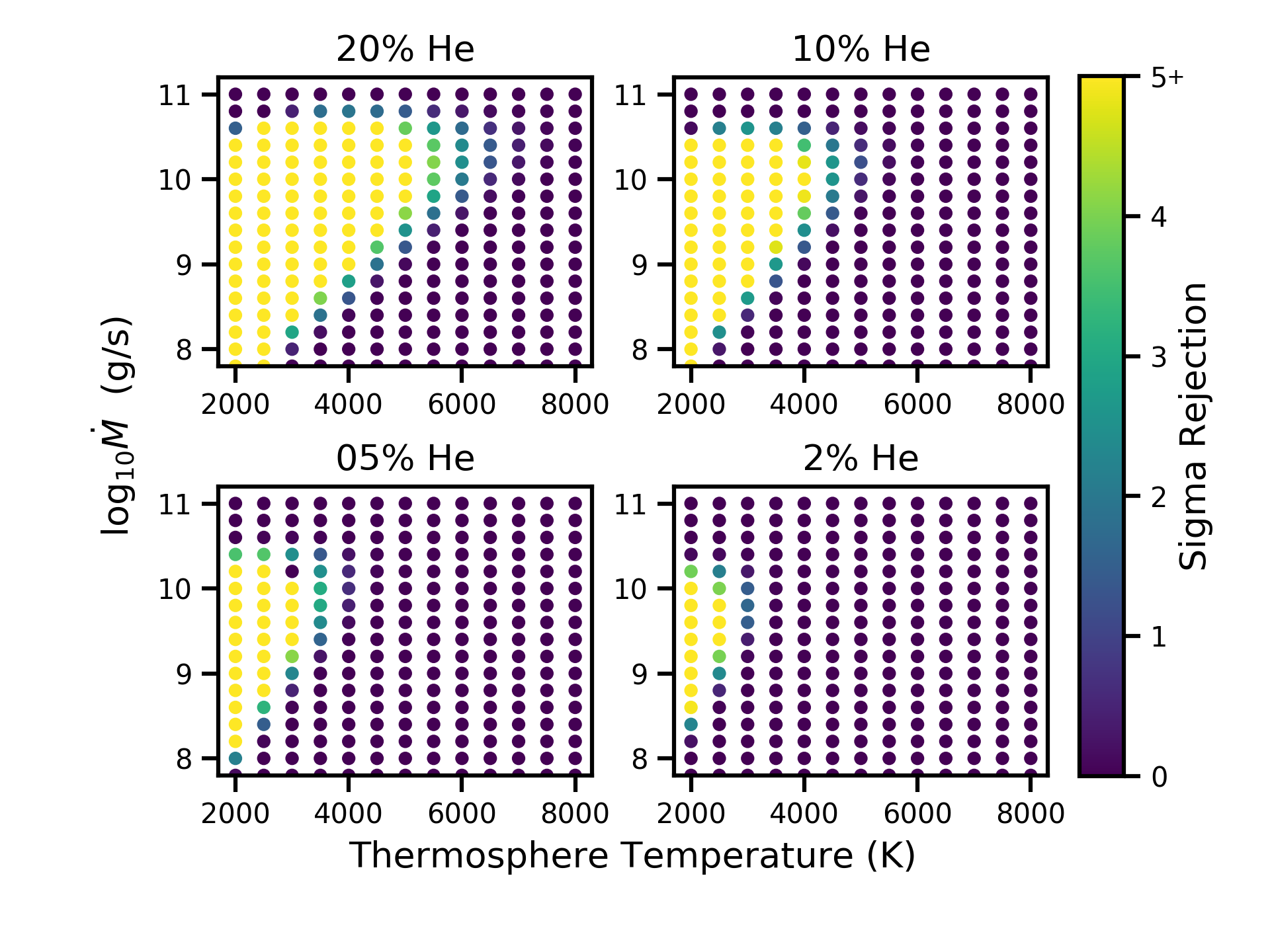}
    \caption{Maps showing the regions of parameter space which are rejected by our data. We compared our absorption spectrum to the helium absorption signal produced by our one-dimensional, escaping atmosphere models with varying thermospheric temperatures and mass-loss rates. The colour indicates the statistical significance of the deviation from a good fit, as described in Section \ref{sec:modelling}. 
    }
    \label{fig:gridmod}
\end{figure*}

To determine which regions of parameter space could be excluded by our data, we first combined the two planetary absorption spectra from orders 70 and 71; broadened each model absorption spectrum to the instrumental resolution (R=19,000); and resampled them onto the wavelengths of the observations. For each model, we found the optimal window around the 10830\AA\, feature by maximizing the cumulative distribution function of the model absorption (defined at each pixel point) over the standard deviation of the data (defined in a constant $\pm$1.5 \AA\, region around the feature). The window was centered on the middle of the two stronger lines of the triplet feature to maximize the contrast. The observed data region to use for testing the models was thus chosen as a compromise between sampling the true spread in the data assuming the null hypothesis and retaining information in the relevant region assuming the contrary. Following the optimization of the test window, the corresponding (maximal) rejection of the model given the data was found by its relation to the cumulative distribution function. Figure \ref{fig:gridmod} shows the map of the statistical deviation from our data for the model grids. We note that this method does not take into account any correlated noise present in the data.

The three parameters varied in our model grids (mass-loss rate; thermospheric temperature; and helium abundance) control the size of the absorption feature, and determine which regions of parameter space can be ruled out under the assumption that the atmosphere can be described as a one-dimensional, isothermal Parker wind. Increasing the mass loss rate increases the absorption signal. In general, decreasing the temperature in the model increases the size of the helium absorption feature. This is because higher temperatures lead to faster outflow velocities, leading to lower densities for a given mass-loss rate; and because lower temperatures lead to lower metastable helium production rates through recombination. Increasing the helium number fraction increases the size of the model absorption feature. Our non-detection therefore rules out more models at higher mass-loss rates, lower temperatures, and higher helium abundances. 
The absorption signals in our model grids were weaker than those calculated by \cite{Kasper_2020}, because we assumed a greater planet mass and a weaker UV spectrum. Additionally, our uncertainties around the telluric absorption lines were larger than theirs because we removed contaminated pixels, whereas \cite{Kasper_2020} corrected contaminated pixels with Molecfit. These two factors meant that our analysis rejected a smaller region of parameter space than theirs did.

We evaluated the magnitude of the disagreement between our measurement and that of \cite{2022A&A...659A..55O} by repeating the same significance-of-rejection calculation using their best-fit Gaussian model. This model has an absorption amplitude of 2.1\%, standard deviation of 0.55\AA, and a peak wavelength of 10830.10\AA. It is rejected by our data at more than 10 $\sigma$ confidence (ignoring correlated noise). To illustrate the difficulty of predicting helium absorption strengths for exoplanets, we also compared our data to a forward model which used the best available predictions for the planetary mass-loss rate and thermospheric temperature for GJ 1214b.  Previously, \cite{Salz2016} calculated detailed radiative-hydrodynamical models for several specific planets, including GJ 1214b. Their models are one-dimensional, and include hydrogen and helium only. For GJ 1214b, they found maximum thermospheric temperatures of around $T$ = 2500 K, and a mass-loss rate of $\dot{M} =$ 5$\times 10^{9}$ g/s. For these suggested parameters, we show the closest-matching model from our grid in Figure \ref{fig:comparison} ($T = 2600K, \dot{M} = 5\times 10^{9}$ g/s). This fiducial model is much narrower than the best-fit Gaussian model from the CARMENES data, and has a maximum absorption of around 3\% in the core of the helium line. It is ruled out by our data at over 5 $\sigma$ confidence. In all of our model grids of varying helium number fractions, mass-loss models at $T$ = 2500 K and $\dot{M} =$ 5$\times 10^{9}$ g/s are ruled out at over 5 $\sigma$ confidence.

\section{Discussion}\label{sec:discussion}

The EUV flux of GJ 1214 has not been measured, and is a significant source of uncertainty in our models. \cite{Oklopcic2019} noted that a factor of ten uncertainty in the input EUV flux is within reason, and a ten-times reduction in EUV flux typically corresponds to a factor-of-three reduction in the size of the helium absorption feature at mid-transit. This could explain much of the inconsistency between the fiducial model and our observations. Indeed, \cite{2022MNRAS.512.1751P} performed a detailed analysis of the effect of the temperature and iron abundances of stellar coronae on their narrowband EUV emission, and found a strong link between the strength of EUV iron emission lines and the exoplanetary helium absorption strength. She suggests that GJ1214b has weak EUV iron emission lines, which would lead to low rates of metastable helium production in the planet's atmosphere, even if it has a large, helium-rich atmosphere.  Alternatively, \cite{Salz2016} note many important factors that are missing from their models: metals cooling the outflow and reducing the mass-loss rate; stellar wind forcing the escaping material radially outwards away from the star (e.g. \citealt{2021ApJ...914...99W, 2022ApJ...926..226M}); and suppression of the planetary outflow by magnetic fields. This last effect was first studied by \cite{2011ApJ...730...27A}, \cite{2011ApJ...728..152T}, and \cite{2014MNRAS.444.3761O}; and was suggested to be occuring on WASP-52b, WASP-80b \citep{2022arXiv220411865V}; and TOI-560b \citep{2022AJ....163...67Z}.  Any combination of the above effects, and a heluim abundance of less than 10\% in the upper atmosphere, may be the reason for our non-detection.

There are several possible explanations for the discrepancy between the Keck and CARMENES datasets: instrumental effects; variability of the stellar flux (leading to differing fractions of helium in the metastable state); variability in the planetary wind (e.g. due to hydrodynamical instabilities); or stochastic, correlated noise. CARMENES and Keck/NIRSPEC have previously been shown to measure consistent exoplanet absorption spectra around 10830 \AA, as in the case of WASP-107b \citep{Kirk2020, Allart2019, 2021AJ....162..284S}, so the discrepancy between the datasets is unlikely to be due to the use of different instruments. There is no evidence of flares or photometric variability for GJ 1214 over the 1\% level in three years of MEarth data \citep{2011ApJ...736...12B}, and 21 days of Spitzer observations \citep{2013ApJ...765..127F}. Our observed stellar spectrum (Figure \ref{fig:invsout}) is also very similar that observed by CARMENES around 10830 \AA. Neither spectrum shows a strong stellar helium line, which forms in the chromosphere and is correlated with stellar activity for cool stars (e.g. \citealt{1968ApJ...152..123V, 2020A&A...640A..52F}). \cite{2022A&A...659A..55O} also measured several activity indicators at the time of their observations (e.g. H$\alpha$, Na I D1 and D2, and
Ca II IRT line indices), and did not find any evidence for strong activity or flares. It is therefore difficult to imagine stellar variability as the culprit. Three-dimensional simulations of planetary outflows which include the effects of stellar wind have shown that hydrodynamical instabilities can cause the shape of the outflow (and hence absorption strengths in transit) to vary in time (e.g. \citealt{2019ApJ...873...89M, 2021ApJ...914...99W}). In their models of the inflated gas-giant WASP-107b, \cite{2021ApJ...914...99W} show that instablities can cause the helium absorption signal to vary by $\pm$10\% of its measured value, over hours-long time scales. To explain the discrepancy between the Keck and CARMENES observations of GJ 1214b (which has a much lower gravitational potential) would require much larger variations, on the order of 100\% of the observed signal measured by \cite{2022A&A...659A..55O}. Detailed simulations of GJ 1214b's outflow, including stellar wind effects, would be helpful to investigate whether this scenario is plausible for such a low-mass planet. Finally, correlated noise of unknown origin can be seen in both datasets in Figure \ref{fig:comparison}, and in Section \ref{sec:absspec} we calculated a systematic error of 0.42\% on the absorption depth measured by Keck. An unfortunate alignment of the correlated noise around 10830 \AA\, in either the Keck or the CARMENES datasets could contribute to some of the observed variations around the helium lines.  

In summary, it is currently difficult to determine whether astrophysical variability or correlated noise is responsible for descrepency between the datasets. However, in July 2022, GJ 1214b was observed with JWST's Mid-Infrared Instrument. These observations may shed light on the size and composition of GJ 1214b's atmosphere. If the planet is shown to have a large hydrogen-helium envelope, then the tentative detection of \cite{2022A&A...659A..55O} will seem more likely to be astrophysical in nature, and GJ 1214b's helium signal will seem more likely to be truly variable. Further helium observations will then be necessary to better characterize the variability. Since our results show that GJ 1214b has a maximum helium absorption depth of 1.27\% at least some of the time - even when the data is uncontaminated by telluric absorption and emission lines - we may perhaps assume that the non-detection of \cite{2020AJ....160..258K} was not caused by telluric contamination. That means there have so far been three published non-detections of helium in GJ 1214b that had the required precision to detect the 2.1\% transit depth observed by \cite{2020AJ....160..258K}. These are: this work; that of \cite{2020AJ....160..258K}; and that of \cite{2020RNAAS...4..231P} (the non-detection of \cite{Crossfield_2019} had a 95\% confidence upper-limit of 2.1 $R_{p}$, or 6.0\%). The observations spanned the years 2011 to 2021. With only one of four observations showing a detectable signature in that time, one may expect an observing campaign (with Keck or CARMENES) targeting four transits over several years to yield one additional helium detection on GJ 1214b. If no further detections are made in that time, then true variability in the planetary signal seems less likely.

\section{Conclusion}\label{sec:conclusion}
Transits of GJ1214b have been observed multiple times at high resolution around the 10830 \AA\,  helium line (\citealt{2022A&A...659A..55O, 2020RNAAS...4..231P, Crossfield_2019, 2020AJ....160..258K}). Most of the observations have been non-detections, except for one transit event, observed with CARMENES, which showed a tentative absorption feature reaching over 2\% \citep{2022A&A...659A..55O}. The authors of the study noted that previous non-detections were contaminated by telluric emission- and absorption lines overlapping the planetary signal. We observed GJ1214b with NIRSPEC on Keck, at a time of minimal telluric contamination, and did not detect the same 2\% absorption signal. The best-fit Gaussian model of \citep{2022A&A...659A..55O} is rejected by our data at over 10 $\sigma$ confidence, when we ignore correlated noise in the spectrum. When correlated noise is taken into account, we find a 95\% confidence upper limit of 1.22\%, or 1.37 R$_{p}$. We explored several possible explanations for the discrepancy between the Keck and CARMENES datasets, and found that it is unlikely to be caused by the use of different instruments, or stellar variability. Whether it can be explained by large variations in the planetary wind itself (e.g. due to hydrodynamical instabilities), or correlated noise, may be revealed by upcoming JWST observations or further ground-based helium measurements.

We also showed that 1D Parker wind models which use thermospheric temperatures and mass-loss rates implied by radiative-hydrodynamical calculations of \cite{Salz2016} are ruled out at over 5 $\sigma$ confidence. Possible explanations include uncertainties in the stellar EUV flux; a lower-than-10\% helium number fraction in the upper atmosphere; metals cooling the upper atmosphere and lowering the mass-loss rate; and 3D effects like stellar wind reducing the absorption signal size; or combinations thereof.

\begin{acknowledgments}
We are deeply grateful to the WMKO support astronomers who provided exceptional support during this observing program, which occurred soon after the NIRSPEC upgrade; in particular, Carlos Alvarez who assisted us with this run, and Greg Doppmann who assisted us with others.
We also wish to recognize and acknowledge the very significant cultural role and reverence that the summit of Maunakea has always had within the indigenous Hawaiian community.  We are most fortunate to have the opportunity to conduct observations from this mountain.
We thank the kind, anonymous referee for their comments on the manuscript.
The data presented herein were obtained at the W. M. Keck Observatory, which is operated as a scientific partnership among the California Institute of Technology, the University of California and the National Aeronautics and Space Administration. The Observatory was made possible by the generous financial support of the W. M. Keck Foundation.

\end{acknowledgments}


\vspace{5mm}
\facility{Keck:II (NIRSPEC)} 
\software{SciPy \citep{2020SciPy-NMeth};  NumPy \citep{van2011numpy};  matplotlib \citep{Hunter:2007}; Astropy \citep{2013A&A...558A..33A}}

%\appendix

%\section{Appendix information}

\bibliography{sample631}{}

\begin{thebibliography}{}
\expandafter\ifx\csname natexlab\endcsname\relax\def\natexlab#1{#1}\fi
\providecommand{\url}[1]{\href{#1}{#1}}
\providecommand{\dodoi}[1]{doi:~\href{http://doi.org/#1}{\nolinkurl{#1}}}
\providecommand{\doeprint}[1]{\href{http://ascl.net/#1}{\nolinkurl{http://ascl.net/#1}}}
\providecommand{\doarXiv}[1]{\href{https://arxiv.org/abs/#1}{\nolinkurl{https://arxiv.org/abs/#1}}}

\bibitem[{{Adams}(2011)}]{2011ApJ...730...27A}
{Adams}, F.~C. 2011, \apj, 730, 27, \dodoi{10.1088/0004-637X/730/1/27}

\bibitem[{{Allart} {et~al.}(2018){Allart}, {Bourrier}, {Lovis}, {Ehrenreich},
  {Spake}, {Wyttenbach}, {Pino}, {Pepe}, {Sing}, \& {Lecavelier des
  Etangs}}]{Allart2018}
{Allart}, R., {Bourrier}, V., {Lovis}, C., {et~al.} 2018, Science, 362, 1384,
  \dodoi{10.1126/science.aat5879}

\bibitem[{{Allart} {et~al.}(2019){Allart}, {Bourrier}, {Lovis}, {Ehrenreich},
  {Aceituno}, {Guijarro}, {Pepe}, {Sing}, {Spake}, \&
  {Wyttenbach}}]{Allart2019}
---. 2019, arXiv e-prints, arXiv:1901.08073.
\newblock \doarXiv{1901.08073}

\bibitem[{{Astropy Collaboration} {et~al.}(2013){Astropy Collaboration},
  {Robitaille}, {Tollerud}, {Greenfield}, {Droettboom}, {Bray}, {Aldcroft},
  {Davis}, {Ginsburg}, {Price-Whelan}, {Kerzendorf}, {Conley}, {Crighton},
  {Barbary}, {Muna}, {Ferguson}, {Grollier}, {Parikh}, {Nair}, {Unther},
  {Deil}, {Woillez}, {Conseil}, {Kramer}, {Turner}, {Singer}, {Fox}, {Weaver},
  {Zabalza}, {Edwards}, {Azalee Bostroem}, {Burke}, {Casey}, {Crawford},
  {Dencheva}, {Ely}, {Jenness}, {Labrie}, {Lim}, {Pierfederici}, {Pontzen},
  {Ptak}, {Refsdal}, {Servillat}, \& {Streicher}}]{2013A&A...558A..33A}
{Astropy Collaboration}, {Robitaille}, T.~P., {Tollerud}, E.~J., {et~al.} 2013,
  \aap, 558, A33, \dodoi{10.1051/0004-6361/201322068}

\bibitem[{{Barrag{\'a}n} {et~al.}(2021){Barrag{\'a}n}, {Armstrong}, {Gandolfi},
  {Carleo}, {Vidotto}, {Villarreal D'Angelo}, {Oklop{\v{c}}i{\'c}}, {Isaacson},
  {Oddo}, {Collins}, {Fridlund}, {Sousa}, {Persson}, {Hellier}, {Howell},
  {Howard}, {Redfield}, {Eisner}, {Georgieva}, {Dragomir}, {Bayliss},
  {Nielsen}, {Klein}, {Aigrain}, {Zhang}, {Teske}, {Twicken}, {Jenkins},
  {Esposito}, {Van Eylen}, {Rodler}, {Adibekyan}, {Alarcon}, {Anderson}, {Akana
  Murphy}, {Barrado}, {Barros}, {Benneke}, {Bouchy}, {Bryant}, {Butler},
  {Burt}, {Cabrera}, {Casewell}, {Chaturvedi}, {Cloutier}, {Cochran}, {Crane},
  {Crossfield}, {Crouzet}, {Collins}, {Dai}, {Deeg}, {Deline}, {Demangeon},
  {Dumusque}, {Figueira}, {Furlan}, {Gnilka}, {Goad}, {Goffo},
  {Guti{\'e}rrez-Canales}, {Hadjigeorghiou}, {Hartman}, {Hatzes}, {Harris},
  {Henderson}, {Hirano}, {Hojjatpanah}, {Hoyer}, {Kab{\'a}th}, {Korth},
  {Lillo-Box}, {Luque}, {Marmier}, {Mo{\v{c}}nik}, {Muresan}, {Murgas},
  {Nagel}, {Osborne}, {Osborn}, {Osborn}, {Palle}, {Raimbault}, {Ricker},
  {Rubenzahl}, {Stockdale}, {Santos}, {Scott}, {Schwarz}, {Shectman},
  {Raimbault}, {Seager}, {S{\'e}gransan}, {Serrano}, {Skarka}, {Smith},
  {{\v{S}}ubjak}, {Tan}, {Udry}, {Watson}, {Wheatley}, {West}, {Winn}, {Wang},
  {Wolfgang}, \& {Ziegler}}]{2021arXiv211013069B}
{Barrag{\'a}n}, O., {Armstrong}, D.~J., {Gandolfi}, D., {et~al.} 2021, arXiv
  e-prints, arXiv:2110.13069.
\newblock \doarXiv{2110.13069}

\bibitem[{{Bean} {et~al.}(2010){Bean}, {Miller-Ricci Kempton}, \&
  {Homeier}}]{2010Natur.468..669B}
{Bean}, J.~L., {Miller-Ricci Kempton}, E., \& {Homeier}, D. 2010, \nat, 468,
  669, \dodoi{10.1038/nature09596}

\bibitem[{{Bean} {et~al.}(2021){Bean}, {Raymond}, \&
  {Owen}}]{2021JGRE..12606639B}
{Bean}, J.~L., {Raymond}, S.~N., \& {Owen}, J.~E. 2021, Journal of Geophysical
  Research (Planets), 126, e06639, \dodoi{10.1029/2020JE006639}

\bibitem[{{Berta} {et~al.}(2011){Berta}, {Charbonneau}, {Bean}, {Irwin},
  {Burke}, {D{\'e}sert}, {Nutzman}, \& {Falco}}]{2011ApJ...736...12B}
{Berta}, Z.~K., {Charbonneau}, D., {Bean}, J., {et~al.} 2011, \apj, 736, 12,
  \dodoi{10.1088/0004-637X/736/1/12}

\bibitem[{{Berta} {et~al.}(2012){Berta}, {Charbonneau}, {D{\'e}sert},
  {Miller-Ricci Kempton}, {McCullough}, {Burke}, {Fortney}, {Irwin}, {Nutzman},
  \& {Homeier}}]{2012ApJ...747...35B}
{Berta}, Z.~K., {Charbonneau}, D., {D{\'e}sert}, J.-M., {et~al.} 2012, \apj,
  747, 35, \dodoi{10.1088/0004-637X/747/1/35}

\bibitem[{{Charbonneau} {et~al.}(2009){Charbonneau}, {Berta}, {Irwin}, {Burke},
  {Nutzman}, {Buchhave}, {Lovis}, {Bonfils}, {Latham}, {Udry}, {Murray-Clay},
  {Holman}, {Falco}, {Winn}, {Queloz}, {Pepe}, {Mayor}, {Delfosse}, \&
  {Forveille}}]{2009Natur.462..891C}
{Charbonneau}, D., {Berta}, Z.~K., {Irwin}, J., {et~al.} 2009, \nat, 462, 891,
  \dodoi{10.1038/nature08679}

\bibitem[{{Cloutier} {et~al.}(2021){Cloutier}, {Charbonneau}, {Deming},
  {Bonfils}, \& {Astudillo-Defru}}]{2021AJ....162..174C}
{Cloutier}, R., {Charbonneau}, D., {Deming}, D., {Bonfils}, X., \&
  {Astudillo-Defru}, N. 2021, \aj, 162, 174, \dodoi{10.3847/1538-3881/ac1584}

\bibitem[{Crossfield {et~al.}(2019)Crossfield, Barman, Hansen, \&
  Frewen}]{Crossfield_2019}
Crossfield, I. J.~M., Barman, T., Hansen, B., \& Frewen, S. 2019, Research
  Notes of the {AAS}, 3, 24, \dodoi{10.3847/2515-5172/ab01b8}

\bibitem[{{Fraine} {et~al.}(2013){Fraine}, {Deming}, {Gillon}, {Jehin},
  {Demory}, {Benneke}, {Seager}, {Lewis}, {Knutson}, \&
  {D{\'e}sert}}]{2013ApJ...765..127F}
{Fraine}, J.~D., {Deming}, D., {Gillon}, M., {et~al.} 2013, \apj, 765, 127,
  \dodoi{10.1088/0004-637X/765/2/127}

\bibitem[{{Fuhrmeister} {et~al.}(2020){Fuhrmeister}, {Czesla}, {Hildebrandt},
  {Nagel}, {Schmitt}, {Jeffers}, {Caballero}, {Hintz}, {Johnson},
  {Sch{\"o}fer}, {Zechmeister}, {Reiners}, {Ribas}, {Amado}, {Quirrenbach},
  {Nortmann}, {Bauer}, {B{\'e}jar}, {Cort{\'e}s-Contreras}, {Dreizler},
  {Galad{\'\i}-Enr{\'\i}quez}, {Hatzes}, {Kaminski}, {K{\"u}rster}, {Lafarga},
  \& {Montes}}]{2020A&A...640A..52F}
{Fuhrmeister}, B., {Czesla}, S., {Hildebrandt}, L., {et~al.} 2020, \aap, 640,
  A52, \dodoi{10.1051/0004-6361/202038279}

\bibitem[{{Fulton} {et~al.}(2017){Fulton}, {Petigura}, {Howard}, {Isaacson},
  {Marcy}, {Cargile}, {Hebb}, {Weiss}, {Johnson}, {Morton}, {Sinukoff},
  {Crossfield}, \& {Hirsch}}]{2017AJ....154..109F}
{Fulton}, B.~J., {Petigura}, E.~A., {Howard}, A.~W., {et~al.} 2017, \aj, 154,
  109, \dodoi{10.3847/1538-3881/aa80eb}

\bibitem[{{Horne}(1986)}]{1986PASP...98..609H}
{Horne}, K. 1986, \pasp, 98, 609, \dodoi{10.1086/131801}

\bibitem[{Hunter(2007)}]{Hunter:2007}
Hunter, J.~D. 2007, Computing In Science \& Engineering, 9, 90

\bibitem[{{Husser} {et~al.}(2013){Husser}, {Wende-von Berg}, {Dreizler},
  {Homeier}, {Reiners}, {Barman}, \& {Hauschildt}}]{2013A&A...553A...6H}
{Husser}, T.~O., {Wende-von Berg}, S., {Dreizler}, S., {et~al.} 2013, \aap,
  553, A6, \dodoi{10.1051/0004-6361/201219058}

\bibitem[{{Kasper} {et~al.}(2020){Kasper}, {Bean}, {Oklop{\v{c}}i{\'c}},
  {Malsky}, {Kempton}, {D{\'e}sert}, {Rogers}, \&
  {Mansfield}}]{2020AJ....160..258K}
{Kasper}, D., {Bean}, J.~L., {Oklop{\v{c}}i{\'c}}, A., {et~al.} 2020, \aj, 160,
  258, \dodoi{10.3847/1538-3881/abbee6}

\bibitem[{Kasper {et~al.}(2020)Kasper, Bean, Oklopčić, Malsky, Kempton,
  Désert, Rogers, \& Mansfield}]{Kasper_2020}
Kasper, D., Bean, J.~L., Oklopčić, A., {et~al.} 2020, The Astronomical
  Journal, 160, 258, \dodoi{10.3847/1538-3881/abbee6}

\bibitem[{{Kim} {et~al.}(2015){Kim}, {Prato}, \&
  {McLean}}]{2015ascl.soft07017K}
{Kim}, S., {Prato}, L., \& {McLean}, I. 2015, {REDSPEC: NIRSPEC data
  reduction}.
\newblock \doeprint{1507.017}

\bibitem[{{Kirk} {et~al.}(2020){Kirk}, {Alam}, {L{\'o}pez-Morales}, \&
  {Zeng}}]{Kirk2020}
{Kirk}, J., {Alam}, M.~K., {L{\'o}pez-Morales}, M., \& {Zeng}, L. 2020, \aj,
  159, 115, \dodoi{10.3847/1538-3881/ab6e66}

\bibitem[{{Kreidberg} \& {Oklop{\v{c}}i{\'c}}(2018)}]{KreidbergOklopcic2018}
{Kreidberg}, L., \& {Oklop{\v{c}}i{\'c}}, A. 2018, Research Notes of the
  American Astronomical Society, 2, 44, \dodoi{10.3847/2515-5172/aac887}

\bibitem[{{Kreidberg} {et~al.}(2014){Kreidberg}, {Bean}, {D{\'e}sert},
  {Benneke}, {Deming}, {Stevenson}, {Seager}, {Berta-Thompson}, {Seifahrt}, \&
  {Homeier}}]{2014Natur.505...69K}
{Kreidberg}, L., {Bean}, J.~L., {D{\'e}sert}, J.-M., {et~al.} 2014, \nat, 505,
  69, \dodoi{10.1038/nature12888}

\bibitem[{{Lalitha} {et~al.}(2014){Lalitha}, {Poppenhaeger}, {Singh}, {Czesla},
  \& {Schmitt}}]{2014ApJ...790L..11L}
{Lalitha}, S., {Poppenhaeger}, K., {Singh}, K.~P., {Czesla}, S., \& {Schmitt},
  J.~H.~M.~M. 2014, \apjl, 790, L11, \dodoi{10.1088/2041-8205/790/1/L11}

\bibitem[{{Lopez} \& {Fortney}(2013)}]{2013ApJ...776....2L}
{Lopez}, E.~D., \& {Fortney}, J.~J. 2013, \apj, 776, 2,
  \dodoi{10.1088/0004-637X/776/1/2}

\bibitem[{{MacLeod} \& {Oklop{\v{c}}i{\'c}}(2022)}]{2022ApJ...926..226M}
{MacLeod}, M., \& {Oklop{\v{c}}i{\'c}}, A. 2022, \apj, 926, 226,
  \dodoi{10.3847/1538-4357/ac46ce}

\bibitem[{Mansfield {et~al.}(2018)Mansfield, Bean, Oklop{\v{c}}i{\'c},
  Kreidberg, Désert, Kempton, Line, Fortney, Henry, Mallonn, Stevenson,
  Dragomir, Allart, \& Bourrier}]{Mansfield2018}
Mansfield, M., Bean, J.~L., Oklop{\v{c}}i{\'c}, A., {et~al.} 2018, ApJL, 868,
  L34

\bibitem[{{Martin} {et~al.}(2018){Martin}, {Fitzgerald}, {McLean}, {Doppmann},
  {Kassis}, {Aliado}, {Canfield}, {Johnson}, {Kress}, {Lanclos}, {Magnone},
  {Sohn}, {Wang}, \& {Weiss}}]{Martin2018}
{Martin}, E.~C., {Fitzgerald}, M.~P., {McLean}, I.~S., {et~al.} 2018, Proc.
  SPIE, 10702, 107020A, \dodoi{10.1117/12.2312266}

\bibitem[{{McCann} {et~al.}(2019){McCann}, {Murray-Clay}, {Kratter}, \&
  {Krumholz}}]{2019ApJ...873...89M}
{McCann}, J., {Murray-Clay}, R.~A., {Kratter}, K., \& {Krumholz}, M.~R. 2019,
  \apj, 873, 89, \dodoi{10.3847/1538-4357/ab05b8}

\bibitem[{{McLean} {et~al.}(2000){McLean}, {Graham}, {Becklin}, {Figer},
  {Larkin}, {Levenson}, \& {Teplitz}}]{mclean2000}
{McLean}, I.~S., {Graham}, J.~R., {Becklin}, E.~E., {et~al.} 2000, Proc. SPIE,
  4008, 1048, \dodoi{10.1117/12.395422}

\bibitem[{{McLean} {et~al.}(1998){McLean}, {Becklin}, {Bendiksen}, {Brims},
  {Canfield}, {Figer}, {Graham}, {Hare}, {Lacayanga}, {Larkin}, {Larson},
  {Levenson}, {Magnone}, {Teplitz}, \& {Wong}}]{mclean1998}
{McLean}, I.~S., {Becklin}, E.~E., {Bendiksen}, O., {et~al.} 1998, Proc. SPIE,
  3354, 566, \dodoi{10.1117/12.317283}

\bibitem[{{Nortmann} {et~al.}(2018){Nortmann}, {Pall{\'e}}, {Salz},
  {Sanz-Forcada}, {Nagel}, {Alonso-Floriano}, {Czesla}, {Yan}, {Chen},
  {Snellen}, {Zechmeister}, {Schmitt}, {L{\'o}pez-Puertas}, {Casasayas-Barris},
  {Bauer}, {Amado}, {Caballero}, {Dreizler}, {Henning}, {Lamp{\'o}n}, {Montes},
  {Molaverdikhani}, {Quirrenbach}, {Reiners}, {Ribas}, {S{\'a}nchez-L{\'o}pez},
  {Schneider}, \& {Zapatero Osorio}}]{Nortmann2018}
{Nortmann}, L., {Pall{\'e}}, E., {Salz}, M., {et~al.} 2018, Science, 362, 1388,
  \dodoi{10.1126/science.aat5348}

\bibitem[{{Oklop{\v{c}}i{\'c}}(2019)}]{Oklopcic2019}
{Oklop{\v{c}}i{\'c}}, A. 2019, \apj, 881, 133, \dodoi{10.3847/1538-4357/ab2f7f}

\bibitem[{{Oklop{\v{c}}i{\'c}} \& {Hirata}(2018)}]{OklopcicHirata2018}
{Oklop{\v{c}}i{\'c}}, A., \& {Hirata}, C.~M. 2018, \apj, 855, L11,
  \dodoi{10.3847/2041-8213/aaada9}

\bibitem[{{Orell-Miquel} {et~al.}(2022){Orell-Miquel}, {Murgas}, {Pall{\'e}},
  {Lamp{\'o}n}, {L{\'o}pez-Puertas}, {Sanz-Forcada}, {Nagel}, {Kaminski},
  {Casasayas-Barris}, {Nortmann}, {Luque}, {Molaverdikhani}, {Sedaghati},
  {Caballero}, {Amado}, {Bergond}, {Czesla}, {Hatzes}, {Henning},
  {Khalafinejad}, {Montes}, {Morello}, {Quirrenbach}, {Reiners}, {Ribas},
  {S{\'a}nchez-L{\'o}pez}, {Schweitzer}, {Stangret}, {Yan}, \& {Zapatero
  Osorio}}]{2022A&A...659A..55O}
{Orell-Miquel}, J., {Murgas}, F., {Pall{\'e}}, E., {et~al.} 2022, \aap, 659,
  A55, \dodoi{10.1051/0004-6361/202142455}

\bibitem[{{Owen} \& {Adams}(2014)}]{2014MNRAS.444.3761O}
{Owen}, J.~E., \& {Adams}, F.~C. 2014, \mnras, 444, 3761,
  \dodoi{10.1093/mnras/stu1684}

\bibitem[{{Owen} \& {Wu}(2013)}]{2013ApJ...775..105O}
{Owen}, J.~E., \& {Wu}, Y. 2013, \apj, 775, 105,
  \dodoi{10.1088/0004-637X/775/2/105}

\bibitem[{Paragas {et~al.}(2021)Paragas, Vissapragada, Knutson, Oklopčić,
  Chachan, Greklek-McKeon, Dai, Tinyanont, \& Vasisht}]{paragas2021metastable}
Paragas, K., Vissapragada, S., Knutson, H.~A., {et~al.} 2021, Metastable Helium
  Reveals an Extended Atmosphere for the Gas Giant HAT-P-18b.
\newblock \doarXiv{2102.08392}

\bibitem[{{Petit dit de la Roche} {et~al.}(2020){Petit dit de la Roche}, {van
  den Ancker}, \& {Miles-Paez}}]{2020RNAAS...4..231P}
{Petit dit de la Roche}, D.~J.~M., {van den Ancker}, M.~E., \& {Miles-Paez},
  P.~A. 2020, Research Notes of the American Astronomical Society, 4, 231,
  \dodoi{10.3847/2515-5172/abcfc7}

\bibitem[{{Poppenhaeger}(2022)}]{2022MNRAS.512.1751P}
{Poppenhaeger}, K. 2022, \mnras, 512, 1751, \dodoi{10.1093/mnras/stac507}

\bibitem[{{Rogers} \& {Seager}(2010)}]{2010ApJ...712..974R}
{Rogers}, L.~A., \& {Seager}, S. 2010, \apj, 712, 974,
  \dodoi{10.1088/0004-637X/712/2/974}

\bibitem[{{Salz} {et~al.}(2016){Salz}, {Czesla}, {Schneider}, \&
  {Schmitt}}]{Salz2016}
{Salz}, M., {Czesla}, S., {Schneider}, P.~C., \& {Schmitt}, J.~H.~M.~M. 2016,
  \aap, 586, A75, \dodoi{10.1051/0004-6361/201526109}

\bibitem[{{Salz} {et~al.}(2018){Salz}, {Czesla}, {Schneider}, {Nagel},
  {Schmitt}, {Nortmann}, {Alonso-Floriano}, {L{\'o}pez-Puertas}, {Lamp{\'o}n},
  {Bauer}, {Snellen}, {Pall{\'e}}, {Caballero}, {Yan}, {Chen}, {Sanz-Forcada},
  {Amado}, {Quirrenbach}, {Ribas}, {Reiners}, {B{\'e}jar}, {Casasayas-Barris},
  {Cort{\'e}s-Contreras}, {Dreizler}, {Guenther}, {Henning}, {Jeffers},
  {Kaminski}, {K{\"u}rster}, {Lafarga}, {Lara}, {Molaverdikhani}, {Montes},
  {Morales}, {S{\'a}nchez-L{\'o}pez}, {Seifert}, {Zapatero Osorio}, \&
  {Zechmeister}}]{Salz2018}
{Salz}, M., {Czesla}, S., {Schneider}, P.~C., {et~al.} 2018, \aap, 620, A97,
  \dodoi{10.1051/0004-6361/201833694}

\bibitem[{{Sanz-Forcada} {et~al.}(2011){Sanz-Forcada}, {Micela}, {Ribas},
  {Pollock}, {Eiroa}, {Velasco}, {Solano}, \&
  {Garc{\'\i}a-{\'A}lvarez}}]{2011A&A...532A...6S}
{Sanz-Forcada}, J., {Micela}, G., {Ribas}, I., {et~al.} 2011, \aap, 532, A6,
  \dodoi{10.1051/0004-6361/201116594}

\bibitem[{{Spake} {et~al.}(2021){Spake}, {Oklop{\v{c}}i{\'c}}, \&
  {Hillenbrand}}]{2021AJ....162..284S}
{Spake}, J.~J., {Oklop{\v{c}}i{\'c}}, A., \& {Hillenbrand}, L.~A. 2021, \aj,
  162, 284, \dodoi{10.3847/1538-3881/ac178a}

\bibitem[{{Spake} {et~al.}(2018){Spake}, {Sing}, {Evans}, {Oklop{\v{c}}i{\'c}},
  {Bourrier}, {Kreidberg}, {Rackham}, {Irwin}, {Ehrenreich}, {Wyttenbach},
  {Wakeford}, {Zhou}, {Chubb}, {Nikolov}, {Goyal}, {Henry}, {Williamson},
  {Blumenthal}, {Anderson}, {Hellier}, {Charbonneau}, {Udry}, \&
  {Madhusudhan}}]{Spake2018}
{Spake}, J.~J., {Sing}, D.~K., {Evans}, T.~M., {et~al.} 2018, \nat, 557, 68,
  \dodoi{10.1038/s41586-018-0067-5}

\bibitem[{{Trammell} {et~al.}(2011){Trammell}, {Arras}, \&
  {Li}}]{2011ApJ...728..152T}
{Trammell}, G.~B., {Arras}, P., \& {Li}, Z.-Y. 2011, \apj, 728, 152,
  \dodoi{10.1088/0004-637X/728/2/152}

\bibitem[{Van Der~Walt {et~al.}(2011)Van Der~Walt, Colbert, \&
  Varoquaux}]{van2011numpy}
Van Der~Walt, S., Colbert, S.~C., \& Varoquaux, G. 2011, Computing in Science
  \& Engineering, 13, 22

\bibitem[{{Vaughan} \& {Zirin}(1968)}]{1968ApJ...152..123V}
{Vaughan}, Arthur~H., J., \& {Zirin}, H. 1968, \apj, 152, 123,
  \dodoi{10.1086/149531}

\bibitem[{{Virtanen} {et~al.}(2020){Virtanen}, {Gommers}, {Oliphant},
  {Haberland}, {Reddy}, {Cournapeau}, {Burovski}, {Peterson}, {Weckesser},
  {Bright}, {van der Walt}, {Brett}, {Wilson}, {Jarrod Millman}, {Mayorov},
  {Nelson}, {Jones}, {Kern}, {Larson}, {Carey}, {Polat}, {Feng}, {Moore}, {Vand
  erPlas}, {Laxalde}, {Perktold}, {Cimrman}, {Henriksen}, {Quintero}, {Harris},
  {Archibald}, {Ribeiro}, {Pedregosa}, {van Mulbregt}, \&
  {Contributors}}]{2020SciPy-NMeth}
{Virtanen}, P., {Gommers}, R., {Oliphant}, T.~E., {et~al.} 2020, Nature
  Methods, \dodoi{https://doi.org/10.1038/s41592-019-0686-2}

\bibitem[{{Vissapragada} {et~al.}(2022){Vissapragada}, {Knutson},
  {Greklek-McKeon}, {Oklopcic}, {Dai}, {dos Santos}, {Jovanovic}, {Mawet},
  {Millar-Blanchaer}, {Paragas}, {Spake}, \& {Vasisht}}]{2022arXiv220411865V}
{Vissapragada}, S., {Knutson}, H.~A., {Greklek-McKeon}, M., {et~al.} 2022,
  arXiv e-prints, arXiv:2204.11865.
\newblock \doarXiv{2204.11865}

\bibitem[{{Wang} \& {Dai}(2021)}]{2021ApJ...914...99W}
{Wang}, L., \& {Dai}, F. 2021, \apj, 914, 99, \dodoi{10.3847/1538-4357/abf1ed}

\bibitem[{{Youngblood} {et~al.}(2016){Youngblood}, {France}, {Loyd}, {Linsky},
  {Redfield}, {Schneider}, {Wood}, {Brown}, {Froning}, {Miguel}, {Rugheimer},
  \& {Walkowicz}}]{2016ApJ...824..101Y}
{Youngblood}, A., {France}, K., {Loyd}, R.~O.~P., {et~al.} 2016, \apj, 824,
  101, \dodoi{10.3847/0004-637X/824/2/101}

\bibitem[{Zeng {et~al.}(2019)Zeng, Jacobsen, Sasselov, Petaev, Vanderburg,
  Lopez-Morales, Perez-Mercader, Mattsson, Li, Heising, Bonomo, Damasso,
  Berger, Cao, Levi, \& Wordsworth}]{zeng2019}
Zeng, L., Jacobsen, S.~B., Sasselov, D.~D., {et~al.} 2019, Proceedings of the
  National Academy of Sciences, 116, 9723–9728,
  \dodoi{10.1073/pnas.1812905116}

\bibitem[{{Zhang} {et~al.}(2022){Zhang}, {Knutson}, {Wang}, {Dai}, \&
  {Barrag{\'a}n}}]{2022AJ....163...67Z}
{Zhang}, M., {Knutson}, H.~A., {Wang}, L., {Dai}, F., \& {Barrag{\'a}n}, O.
  2022, \aj, 163, 67, \dodoi{10.3847/1538-3881/ac3fa7}

\bibitem[{{Zhang} {et~al.}(2020){Zhang}, {Knutson}, {Wang}, {Dai}, {Oklopcic},
  \& {Hu}}]{2020arXiv201202198Z}
{Zhang}, M., {Knutson}, H.~A., {Wang}, L., {et~al.} 2020, arXiv e-prints,
  arXiv:2012.02198.
\newblock \doarXiv{2012.02198}

\end{thebibliography}
\bibliographystyle{aasjournal}



\end{document}